\documentclass[10pt,conference]{IEEEtran}

\usepackage[utf8]{inputenc} 
\usepackage[T1]{fontenc}    
\usepackage{hyperref}       
\usepackage{url}            
\usepackage{booktabs}       
\usepackage{amsfonts}       
\usepackage{nicefrac}       
\usepackage{microtype}      
\usepackage{graphicx}
\usepackage{doi}
\usepackage{wasysym}
\usepackage{pifont}
\usepackage{makecell}
\usepackage{color, colortbl}
\usepackage{paralist}
\usepackage{placeins}
\usepackage{listings}
\usepackage{enumitem}
\usepackage{makecell}
\usepackage{float}
\usepackage{csquotes}
\usepackage{array}

\newcolumntype{x}[1]{>{\centering\arraybackslash\hspace{0pt}}p{#1}}

\usepackage[para,online,flushleft]{threeparttable}
\usepackage{xcolor, tcolorbox}
\usepackage{bold-extra}
\usepackage[style=ieee, backend=biber, maxbibnames=3,natbib=true,isbn=false,doi=false,url=false,eprint=false]{biblatex}

\definecolor{codegreen}{rgb}{0,0.6,0}
\definecolor{codegray}{rgb}{0.5,0.5,0.5}
\definecolor{codepurple}{rgb}{0.58,0,0.82}
\definecolor{backcolour}{rgb}{0.98,0.98,0.98}
\definecolor{bordercolour}{rgb}{0.95,0.95,0.95}
\definecolor{white}{rgb}{1,1,1}

\lstdefinestyle{mystyle}{
	commentstyle=\color{codegreen},
	keywordstyle=\color{magenta},
	numberstyle=\tiny\color{codegray},
	stringstyle=\color{codepurple},
	rulecolor=\color{bordercolour},
	basicstyle=\ttfamily\footnotesize,
	breakatwhitespace=false,         
	breaklines=true,                 
	captionpos=b,                    
	keepspaces=true,                 
	numbers=left,               
	basicstyle=\ttfamily\scriptsize,     
	numbersep=5pt,                  
	showspaces=false,                
	showstringspaces=false,
	showtabs=false,
}


\definecolor{Gray}{gray}{0.95}

\newcommand{\cmark}{\ding{51}}%
\newcommand{\xmark}{\ding{55}}%

\definecolor{Gray}{gray}{0.98}

\newcommand{\resultbox}[1]{
	\begin{tcolorbox}[colback={Gray},colframe=gray!30!white,top=3pt, left=3pt,right=3pt, bottom=3pt]
		#1
	\end{tcolorbox}
}

\title{Automatic Program Repair with OpenAI's Codex\\\LARGE{Evaluating QuixBugs}}

\author{\IEEEauthorblockN{Julian Aron Prenner}
\IEEEauthorblockA{Free University of Bozen-Bolzano\\
\texttt{prenner@inf.unibz.it}}
\and
\IEEEauthorblockN{Romain Robbes}
\IEEEauthorblockA{Free University of Bozen-Bolzano\\
\texttt{rrobbes@unibz.it}}}

\begin{document}
\maketitle

\begin{abstract}
	OpenAI's Codex, a GPT-3 like model trained on a large code corpus, has made headlines in and outside of academia.
	Given a short user-provided description, it is capable of synthesizing code snippets
	that are syntactically and semantically valid in most cases.
	In this work, we want to investigate whether Codex is able to localize
	and fix bugs, a task of central interest in the field of \emph{automated program repair}.
	Our initial evaluation uses the multi-language QuixBugs benchmark (40 bugs in both Python and Java). We find that, despite \emph{not being trained for APR}, Codex is surprisingly effective, and competitive with recent state of the art techniques.
	Our results also show that Codex is slightly more successful at repairing
	Python than Java.
\end{abstract}


\section{Introduction}

Finding and fixing bugs costs billions yearly~\cite{brittonReversibleDebuggingSoftware}
and takes up a considerable proportion of developer time~\cite{latozaMaintainingMentalModels2006}.
The field of Automatic program repair (APR) attempts to develop tools that can automatically find and fix bugs in software. Many existing APR tools follow a test-driven approach: bugs need to be exposed by a failing test case. A variety of different APR approaches have been proposed in the recent years:
\begin{inparaenum}[i)]
	\item using genetic-programming (e.g., GenProg~\citep{gouesGenProgGenericMethod2012} or ARJA~\citep{yuanARJAAutomatedRepair2020}),
	\item using repair patterns (such as PAR~\citep{kimAutomaticPatchGeneration2013}, ELIXIR~\citep{sahaElixirEffectiveObjectoriented2017} or TBar~\citep{liuTBarRevisitingTemplatebased2019}),
	\item code retrieval-based approaches (e.g., ssFix~\citep{xinLeveragingSyntaxrelatedCode2017} or LSRepair~\citep{liuLSRepairLiveSearch2018}),
	\item or using deep learning (e.g., SequenceR~\citep{chenSequenceRSequencetoSequenceLearning2019}, HOPPITY~\citep{dinellaHOPPITYLEARNINGGRAPH2020}, CoCoNuT~\citep{lutellierCoCoNuTCombiningContextaware2020} or CURE~\citep{jiangCURECodeAwareNeural2021}).
\end{inparaenum}


While there has been some promising work using deep neural networks for program repair, several avenues of research are yet to explore in this area. 
In particular, \citeauthor{kaplan2020scaling} observed that Transformer-based language models are subject to several \emph{scaling laws}, including that the performance of a language model has a power law relationship with model size, dataset size, and amount of computing power invested in training the language model, as long as none of these factors is a bottleneck. Other laws show that model performance during training is strongly correlated with out of distribution performance, and that larger models require \emph{less} optimization steps and data points than smaller models to achieve the same performance. Taken together, these laws provide evidence for training very large language models. 

One such model is GPT-3, an auto-regressive Transformer language model with 175 billion parameters, which has set a new state of the art in
many human language understanding tasks~\citep{brownLanguageModelsAre2020}. Unlike previous models (such as BERT \cite{devlinBERTPretrainingDeep2019}) that are pre-trained on unlabelled text and fine-tuned on a target task, GPT-3's size allows it to achieve this performance \emph{without fine-tuning on the target task}, solely through its pre-training as a language model (which consists in ``guessing the next word'' on a large amount of text). Adapting GPT-3 to a particular task is done in a few-shot setting by feeding a task description and a handful to a few dozens examples of the task to the model at inference time, and asking the model to complete the text. In some cases, just feeding the task description without examples (zero-shot setting) shows very good performance. Thus, instead of gathering new data and fine-tuning the model on it, the user's task shifts to defining a prompt that triggers the desired behaviour in the language model. 

Recently, OpenAI\footnote{the authors are not affiliated with OpenAI} released Codex~\citep{chenEvaluatingLargeLanguage2021}, a GPT-3 like
model targeted towards code tasks. Codex is at the core of
CodePilot\footnote{\url{https://copilot.github.com/}}, GitHub's AI coding assistant that provides code completion in Visual Studio Code. In OpenAI demos, Codex is able to synthesize whole functions from a short description.
Codex is mostly used in a zero-shot setting: the input is comprised of a short task description and a final \textit{prompt}.
Codex then generates code that \enquote{naturally} \enquote{completes} the prompt.


In this paper we investigate whether Codex can be applied to the challenging task of Automatic Program Repair. Rather than generating code from scratch, we ask: 1) whether Codex shows promise in repairing buggy code, a task that it was \emph{not trained on}, and 2) which types of prompts yield the best performance (we experiment with 5 different configurations).
Since, unlike the majority of APR tools, Codex supports multiple programming languages, we evaluate performance on two programming languages. This multi-lingual requirement leads us to choose the QuixBugs~\citep{linQuixBugsMultilingualProgram2017} program repair benchmark to conduct this initial investigation in Codex's performance for APR.
QuixBugs contains buggy Python and Java implementations of 40 classical computer
science algorithms, such as counting bits in an integer, calculating the
Levenshtein distance or finding the shortest path in a graph
(see Table~\ref{tab:results} for the full list of algorithms).


We further discuss most common model mistakes and compare repair performance between Python and Java and
between Codex and previous neural repair approaches. We find that, especially considering it was not trained on the task, Codex is surprisingly competitive with three recent APR tools (CURE \cite{jiangCURECodeAwareNeural2021} and DeepDebug \cite{drainDeepDebugFixingPython2021} released in 2021, CoCoNut \cite{lutellierCoCoNuTCombiningContextaware2020}, released in 2020), and is in the lead in Python. We have publicly released all our inputs and Codex' outputs\footnote{\url{https://sandbox.zenodo.org/record/934361}}.

\section{Some Context on Codex}
Codex \cite{chenEvaluatingLargeLanguage2021} is a large deep learning-based language model developed by OpenAI. Like GPT-3, which Codex is based on, it builds on the Transformer \cite{vaswaniAttentionAllYou2017}, a successful neural network architecture, but, following the scaling laws, at a very large scale. Codex' size and the amount of data used to train it are unprecedented in Software Engineering: it has 12 billion parameters and was trained on 54 million GitHub repositories. Being a language model, Codex was trained to complete (partial) input in a meaningful way. Codex supports a variety of programming languages including Python, Java, JavaScript, TypeScript, and others. Compared to GPT-3, Codex is smaller, but has a larger input window (4096 vs 2048 tokens), in order to generate code from a larger context. 

Similarly to GPT-3, Codex is versatile: it is capable of carrying out several different tasks, so long as the tasks can be framed as a \enquote{completion task}, in a zero-shot or few-shot setting. Thus, instead of providing data and re-training the model on it, a user of Codex engages in \emph{prompt engineering}: finding out which prompt (possibly with examples) yields the best performance for the task. It's worth noting that this shift from fine-tuning to prompt engineering is welcome, if only for the reason that fine-tuning the model is out of reach for the vast majority of people due to its sheer size. Rather than running on a user's machine, the model runs in the cloud.

Examples of tasks where Codex can be applied are: intelligent code completion, where the input is code that should be completed (a version of Codex powers GitHub Copilot, a code completion plugin for Visual Studio Code); function synthesis, where the input is a function name and a documentation comment, and the output is a code snippet; code explanation, where the input is a code snippet, and the output is a comment commment; and programming language translation, where the input is a code snippet, followed by a comment indicating the language to translate to.

%


\section{Methodology}

\subsection{Dataset}

In all of our experiments we rely on QuixBugs~\cite{linQuixBugsMultilingualProgram2017},
a benchmark of 40 buggy algorithm implementations, including their correct versions and test cases.
The benchmark is bilingual, all algorithms are implemented in Python and Java.
Unlike the Python versions, which contain docstrings with short descriptions of the respective algorithms, Java versions come without descriptive comments; see Figure~\ref{fig:gcd} for an example.

\begin{figure}[!htbp]
	\centering
	\begin{tabular}{c}
		\begin{lstlisting}[language=Python, numbers=none]
def gcd(a, b):
    if b == 0:
        return a
    else:
       return gcd(a % b, b)
"""
Input:
    a: A nonnegative int
    b: A nonnegative int
	
Greatest Common Divisor

Precondition:
    isinstance(a, int) and
    isinstance(b, int)
Output:
    The greatest int that
    divides evenly into a and b
Example:
    >>> gcd(35, 21)
    7
"""
\end{lstlisting}\\
	\hline
	\begin{lstlisting}[language=Java, numbers=none]
package java_programs;
import java.util.*;

public class GCD {
  public static int gcd(int a, int b) {
    if (b == 0) {
      return a;
    } else {
       return gcd(a % b, b);
    }
  }
}
	\end{lstlisting}\\
\end{tabular}
	\caption{Buggy versions of the algorithm calculating the greatest common divisor between two integers in Python and Java. The recursive call should read \texttt{gcd(b, a \% b)}.}
	\label{fig:gcd}
\end{figure}	

\subsection{Prompt Engineering}

Since the only way to interact with Codex via the prompt that it is given, we experiment with several different input configurations. All configurations use a variation of the following template (Python version):  
\begin{lstlisting}[language=Python, numbers=none]
### fix the bug in the following function
(*\textsl{<buggy function and/or docstring here>}*)
			
### fixed function
\end{lstlisting}

\paragraph*{Code only} We simply input the buggy function (without the docstring in Python) in the template.

\paragraph*{Code with hint} To simulate a more precise bug localization, in this configuration we add hint comments. Specifically, we put the comment \texttt{\#\#\# bug is here} (or \texttt{// bug is here} for Java) before any buggy code line.

\paragraph*{Code with docstring (Python)}  We input the buggy code along with the original docstring, which describes the correct behavior of the function. Note that some docstrings contain examples. For Java, docstrings are not available.

\paragraph*{Docstring only (Python)} This corresponds to the default usage of Codex, in which it synthesizes a full algorithm implementation, and thus acts as a baseline.

\paragraph*{Correct code (Python)} To see if Codex would break already correct code, in this configuration we input the bug-free ground-truth program, instead of the buggy one; ideally, Codex would repeat the code unchanged. We slightly alter the input format changing the first line to \texttt{\#\#\# fix \textbf{a possible bug} in the following function}, indicating that the input \emph{might} be bug-free. 

\paragraph*{Input-output examples} For seven Python bugs that no configuration involving
code (i.e., excluding the docstring-only configuration) could correctly fix, we additionally
tried including input-output examples derived from the corresponding test cases as an additional specification.The input-output examples are given in a docstring-like comment and follow the general format; for instance, the docstring in Figure~\ref{fig:gcd}, contains an example. We include all test cases associated to a program, except those exceeding a certain size (120 characters).

\subsection{Codex Parameters}

\begin{table}[htbp]
	\centering
	\footnotesize
	\caption{Used Codex parameters.}
	\begin{tabular}{lc}
		\toprule
		\textbf{Parameter} & \textbf{Value} \\
		\midrule
		Engine & davinci-codex \\
		Temperature & 0 \\
		Max Tokens & 1024 \\
		Top-p & 1.0 \\
		Frequency Penalty & 0.0 \\
		Presence Penalty & 0.0 \\
		Stop & '\#\#\#', '///' \\
		\bottomrule
	\end{tabular}
	\label{tab:params}
\end{table}

We set Codex' parameters as shown in Table~\ref{tab:params}.
We did not yet systematically investigate the effect of these parameters on repair success, as our evaluation involves a significant manual step.
The temperature and top-p parameters control the randomness of the model: a higher temperature or a lower top-p yield more diverse output. The Codex documentation recommends to set temperature to zero or a low value and not to vary both, temperature and top-p. We set the temperature to zero and top-p to one, which lowers diversity but ought to give
high robustness. The frequency and presence penalties prevent the model from outputing the same tokens repeatedly. Since large portions of the input (everything except the buggy line) should in fact be repeated, we do not use such a penalty.

\subsection{Evaluation}

For each configuration and language we manually evaluated the output of Codex, using the following procedure:
\begin{itemize}
\item When Codex output multiple functions we only considered the first and discarded the remaining output.
\item If the output exactly matched the correct ground-truth patch, we considered it correct.
\item We inspected the code to determine whether the output was semantically equivalent to
the ground truth.
\item When in doubt, we ran the associated test cases.
\item If any test failed, the output was considered incorrect.
\end{itemize}

We observed output that passed all test cases but was semantically incorrect
only for the \texttt{kheapsort} bug, where QuixBugs' test cases do not check
an edge case: the tests simply check for sortedness of the program output;
however, it should only be sorted up to the $k$\textsuperscript{th} largest element.

\emph{Acceptable variations.} Since Codex is generating code as a language model, and is note explicitly trained for program repair, we were more lenient in a few cases. In particular, it is natural for a source code language model to try to avoid defining multiple functions or methods with the same name. Thus, if the output of Codex was a function or method with a slightly different name, but otherwise correct, this was not considered an error; we assume that post-processing could ensure such a rename is done automatically. On the other hand, in several cases, Codex simply repeated the input program (bug included); this was, of course, deemed incorrect output.

For various graph related problems (e.g., \texttt{breath-first-search} or \texttt{detect-cycle}) Codex assumed that the attribute pointing to the next node should be named \texttt{next}, while tests assumed it to be named \texttt{successor}. 
We considered this as semantically correct, as this is a reasonable assumption and the proper name was not specified in the input.

\section{Results}

\subsection{Overall Performance}
Table~\ref{tab:comparison} compares Codex's performance with recent previous work. 
We report results from the literature from three recent neural APR approaches: 
CoCoNut \cite{lutellierCoCoNuTCombiningContextaware2020} uses the Neural Machine Translation (NMT) paradigm of program repair, with an ensemble of CNNs, and supports multiple languages.  DeepDebug \cite{drainDeepDebugFixingPython2021} is a large pre-trained Transformer that also uses the NMT paradigm (Python only): the model is fine-tuned on artificially generated bug-introducing commits, and also stack traces and program context. CURE builds on CoCoNut, adding a pre-trained language model (on Java Code) and filters suggestions based on additional context from static analysis \cite{jiangCURECodeAwareNeural2021}. Note that the assessment for correctness was done manually and evaluation criteria might slightly differ between these works.

Codex correctly fixed considerably more Python than Java bugs (up to 23 Python bugs, only 14 Java bugs), indicating that it can handle Python much better than Java; OpenAI does state that Codex is more capable in Python than other languages. Moreover, it is intriguing to see that Codex, \emph{without explicit training on the task}, outperforms CoCoNuT and DeepDebug on Python, and outperforms CoCoNut in Java. While CURE does outperform Codex in Java, Codex outperforms the only other multi-lingual APR tool (CoCoNut). 

\resultbox{Codex is surprisingly competitive with recent work, and its performance is considerably better for Python than Java.}

\subsection{Performance of different prompts}
Table~\ref{tab:results} provides detailed results for each bug and configuration. For many bugs, the choice of prompt matters significantly. In fact, only 6 bugs are fixed in all scenarios and all languages. On the other hand, the prompts do complement each other: only 8 bugs are not fixed by any of the prompts. 

\paragraph*{Hints are not effective}
Providing a hint comment for precise fault localization was overall not effective in our experiments. For both Python and Java, some bugs were fixed only with hints, but for others adding the hint was harmful. Overall, for Java the total number of fixed bugs was the same, while it decreased from 21 to 19 in Python. 
However, hints led Codex to correctly repair two bugs that could not
be successfully repaired otherwise (\texttt{shunting-yard} and \texttt{minimum-spanning-tree}).

\paragraph*{Synthesis from docstrings}
Table~\ref{tab:results} shows that only providing the Python docstrings, Codex
is able to synthesize a correct solution for 45\% of the problems in QuixBugs. Providing buggy code as a starting point and asking to model to fix it led to five more (+28\%) correct program implementations.

\paragraph*{Additional input-output examples} For the seven bugs that no other configuration could solve, a single one (\texttt{subsequences}) was successfully repaired when adding input-output examples from test cases.

\paragraph*{Correct code}
When requested to fix \emph{a possible bug} in correct code, Codex broke six of the total 40 programs.
In two cases, Codex slightly altered the input program, preserving correctness, however.

\resultbox{Prompts have a major effect on Codex' bug fixing ability}

\begin{table}
	\centering
	\caption{Comparison with previous work. Number of correctly fixed bugs for the Java and Python version of QuixBugs (out of 40).}
	\begin{threeparttable}
	\begin{tabular}{lcc}
		 \toprule
		 & \textbf{Java} & \textbf{Python} \\
		 \midrule
		 DeepDebug~\citep{drainDeepDebugFixingPython2021} & -- & 21 \\
		 CoCoNuT~\citep{lutellierCoCoNuTCombiningContextaware2020} &  13 & 19 \\
		 CURE~\citep{jiangCURECodeAwareNeural2021} &  26 & -- \\
		 Codex &  14 & 23/21\tnote{1} \\
		 \bottomrule
	\end{tabular}
	\begin{tablenotes}
		\item[1] with and without docstring, respectively
	\end{tablenotes}
	\end{threeparttable}
	\label{tab:comparison}
\end{table}

\begin{table}[!htbp]
	\centering
	\scriptsize
	\caption{List of bugs that Codex could fix successfully (\cmark). }
	\setlength\extrarowheight{1pt}
	\begin{tabular}{x{2cm}cccc@{\hspace{0.2cm}}cc}
		\toprule
		& \multicolumn{4}{c}{Python} & \multicolumn{2}{c}{Java} \\
		\cline{2-5}
		\cline{6-7}
		& \thead{code +\\ docstr.} & \thead{code\\only} & \thead{code +\\ hint} & \thead{docstr.\\only} & \thead{code} & \thead{code +\\ hint} \\
		\midrule
		\rowcolor{Gray}
bitcount & \cmark & \cmark & \cmark & \cmark & \cmark & \cmark\\
breadth-first-search & \xmark & \xmark & \xmark & \cmark & \xmark & \xmark\\
\rowcolor{Gray}
bucketsort & \cmark & \cmark & \cmark & \cmark & \cmark & \cmark\\
depth-first-search & \cmark & \cmark & \cmark & \cmark & \xmark & \cmark\\
\rowcolor{Gray}
detect-cycle & \xmark & \xmark & \xmark & \cmark & \xmark & \xmark\\
find-first-in-sorted & \cmark & \cmark & \xmark & \xmark & \cmark & \xmark\\
\rowcolor{Gray}
find-in-sorted & \xmark & \xmark & \xmark & \xmark & \xmark & \xmark\\
flatten & \cmark & \cmark & \cmark & \cmark & \xmark & \xmark\\
\rowcolor{Gray}
gcd & \cmark & \cmark & \cmark & \xmark & \cmark & \cmark\\
get-factors & \cmark & \cmark & \cmark & \cmark & \xmark & \xmark\\
\rowcolor{Gray}
hanoi & \cmark & \cmark & \xmark & \xmark & \xmark & \xmark\\
is-valid-parenthesization & \cmark & \cmark & \cmark & \cmark & \cmark & \cmark\\
\rowcolor{Gray}
kheapsort & \xmark & \cmark & \cmark & \xmark & \xmark & \xmark\\
knapsack & \xmark & \cmark & \xmark & \xmark & \cmark & \cmark\\
\rowcolor{Gray}
kth & \cmark & \xmark & \xmark & \xmark & \cmark & \cmark\\
lcs-length & \cmark & \xmark & \xmark & \xmark & \xmark & \xmark\\
\rowcolor{Gray}
levenshtein & \cmark & \xmark & \xmark & \cmark & \cmark & \cmark\\
lis & \xmark & \xmark & \xmark & \xmark & \xmark & \xmark\\
\rowcolor{Gray}
longest-common-subsequence & \cmark & \cmark & \cmark & \cmark & \cmark & \cmark\\
max-sublist-sum & \cmark & \cmark & \cmark & \cmark & \cmark & \cmark\\
\rowcolor{Gray}
mergesort & \xmark & \xmark & \xmark & \xmark & \cmark & \cmark\\
minimum-spanning-tree & \xmark & \xmark & \cmark & \xmark & \xmark & \xmark\\
\rowcolor{Gray}
next-palindrome & \xmark & \xmark & \xmark & \xmark & \xmark & \xmark\\
next-permutation & \cmark & \xmark & \xmark & \cmark & \xmark & \xmark\\
\rowcolor{Gray}
pascal & \xmark & \xmark & \xmark & \cmark & \xmark & \xmark\\
possible-change & \xmark & \cmark & \xmark & \xmark & \xmark & \xmark\\
\rowcolor{Gray}
powerset & \cmark & \cmark & \cmark & \cmark & \xmark & \xmark\\
quicksort & \cmark & \cmark & \cmark & \cmark & \xmark & \xmark\\
\rowcolor{Gray}
reverse-linked-list & \cmark & \cmark & \cmark & \cmark & \cmark & \cmark\\
rpn-eval & \xmark & \xmark & \xmark & \xmark & \xmark & \xmark\\
\rowcolor{Gray}
shortest-path-length & \xmark & \xmark & \xmark & \xmark & \xmark & \xmark\\
shortest-path-lengths & \cmark & \xmark & \xmark & \xmark & \xmark & \xmark\\
\rowcolor{Gray}
shortest-paths & \xmark & \xmark & \xmark & \xmark & \xmark & \xmark\\
shunting-yard & \xmark & \xmark & \cmark & \xmark & \xmark & \xmark\\
\rowcolor{Gray}
sieve & \cmark & \cmark & \cmark & \xmark & \xmark & \cmark\\
sqrt & \cmark & \cmark & \cmark & \cmark & \cmark & \xmark\\
\rowcolor{Gray}
subsequences & \xmark & \xmark & \xmark & \cmark & \xmark & \xmark\\
to-base & \cmark & \cmark & \cmark & \xmark & \xmark & \xmark\\
\rowcolor{Gray}
topological-ordering & \xmark & \xmark & \xmark & \xmark & \xmark & \xmark\\
wrap & \cmark & \cmark & \cmark & \xmark & \cmark & \cmark\\
\bottomrule
\textbf{Total} & 23 & 21 & 19 & 18 & 14 & 14\\

	\end{tabular}
	\label{tab:results}
\end{table}

\section{Limitations and Future Work}
Codex is a very large language model, that has shown impressive ability in completing source code. In this work, we have evaluated---and found surprisingly competitive---the performance of Codex as an APR tool, \emph{with no further training on the task}.
While this initial evaluation of Codex as an APR tool is promising, it has various limitations.

\paragraph*{More annotators} The correctness of the code output
was assessed by a single annotator. Not only is manual evaluation subjective, it is also prone to mistakes. We hope that we will be able to provide a more reliable evaluation in the future, involving at least two annotators.

\paragraph*{Additional benchmarks and languages}
Currently our evaluation is limited to a single benchmark and two programming languages. Extending this study to benchmarks that involve more complex codebases (e.g., Defects4J~\citep{justDefects4JDatabaseExisting2014} or additional programming languages (e.g., BugsJS~\citep{gyimesiBugsJSBenchmarkJavaScript2019}, a JavaScript benchmark) would provide additional interesting insights into Codex' repair capabilities.

\paragraph*{Data leakage}
Codex was trained on very large amounts of code; only OpenAI staff can know with certainty which repositories were included.
We cannot rule out that the correct ground-truth programs were in Codex' training set. This issue is very difficult to address. There are however mitigating factors:
First, if present, these versions would constitute a tiny portion of the training data (54 million repositories). Second, if the correct program versions were in the training set, so would, very likely, also be the incorrect versions, without labels of which version is correct or incorrect. Moreover, a preliminary study of Codex for GitHub Copilot, found that while the model can indeed repeat data from the training set, this was rare (less than 0.1\% of the cases), concerned code that was cloned many times, and happened mostly when the context was nearly empty \cite{zieglerRecitation}.
Finally, Codex was never specifically trained for the task of repairing or localizing bugs.

\paragraph*{More Automation} For this study, we had to perform several manual steps to validate the correctness of the proposals. This includes removing extraneous output from Codex, and making sure the function/method name was the one expected by the tests. Automating these tasks would make the process significantly smoother.

\paragraph*{Testing multiple outputs} Given the lack of automation, we tried a single completion from Codex for each problem and prompt. With more automation, we would be able to try multiple outputs from Codex (by increasing the temperature). Evaluating more than one output significantly increased the performance of Codex for program synthesis (from 29 up to 70\% \cite{chenEvaluatingLargeLanguage2021}), so this could help APR as well. 

\paragraph*{Fine-Tuning} While the most common use case for Codex is to use it directly after pre-training, a fine-tuning API is available for GPT-3. If such an API is made available for Codex, this would be worthwhile exploring to improve performance.



\section{Acknowledgments}
We would like to thank OpenAI for providing access to their Codex model.

\printbibliography{}

\end{document}